\documentclass{article}
\usepackage[utf8]{inputenc}
\usepackage[a4paper, portrait, margin=0.85in]{geometry}
\usepackage[section]{placeins}
\usepackage{amsmath}
\usepackage{amssymb}
\usepackage{xcolor}
\usepackage{graphicx}
\usepackage[colorlinks,citecolor=blue,urlcolor=blue,bookmarks=false,hypertexnames=true]{hyperref}

\title{Construction of Quantum Target Space from  World-Sheet States using Quantum  State Tomography}

\author{Salman Sajad Wani$^1$, Arshid Shabir$^2$, Junaid Ul Hassan$^2$,\\ S. Kannan$^3$, Hrishikesh Patel$^4$,
C. Sudheesh$^5$, Mir Faizal$^{1,6,7}$
\\\\
\textit{\small $^{1}$Canadian Quantum Research Center, 204-3002, 32 Ave Vernon, BC V1T 2L7 Canada}
\\
\textit{\small $^{2}$Cluster University of Srinagar, Jammu \& Kashmir, India, 190008}
\\
\textit{\small $^3$ISRO Inertial Systems Unit, Thiruvananthapuram, 695 013, India.}
\\
\textit{\small $^4$Department of Physics and Astronomy, University of British Columbia,} \\ \textit{\small 6224  Agricultural Road,
Vancouver, V6T 1Z1, Canada}
\\
\textit{\small $^5$Department of Physics, Indian Institute of Space Science and Technology,} \\ \textit{ \small Thiruvananthapuram, India 695 547}
\\
\textit{\small $^6$Department of Physics and Astronomy,
University of Lethbridge,} \\ \textit{\small Lethbridge, AB T1K 3M4, Canada}
\\
\textit{\small $^7$Irving K. Barber School of Arts and Sciences, University of British Columbia}\\
\textit{\small Okanagan Campus, Kelowna, V1V1V7, Canada}
}
\date{}
\begin{document}

\maketitle
\begin{abstract}
In this paper, we will construct the quantum states of target space coordinates from world-sheet states, using quantum state tomography.  To perform  quantum state tomography of an open string, we will construct  suitable  quadrature operators.  We do this by first defining the quadrature operators in world-sheet, and then using them  to construct the quantum  target space quadrature operators for an open string.  We will  connect the  quantum target space to  classical geometry using  coherent string states. We will be using a novel construction based on a string displacement operator to construct these coherent states. The coherent states of the world-sheet will also be used to construct the coherent states in target space. 
 \end{abstract}
\section{Introduction}
Even though there is a clear  connection between target space and world-sheet in classical theory, it is difficult to find a quantum analog of this connection. However, it  is also  possible to establish a connection between quantum states of world-sheet and the classical target space on which these quantum states propagate. This is done by using   the  renormalization group flow of world-sheet perturbations.  For  the quantum theory of world-sheet to remain   conformally invariant, it is required that the  world-sheet $\beta$-functions vanish. The equations obtained from this procedure are then identified with the  equations of motion for the gravitational  action in the target space. Higher curvature  terms  can be obtained from higher loop corrections  on the world-sheet. So, with this procedure,  a consistent classical target space is constructed on which is consistent with the  quantum theory of world-sheet.  However, it is important to construct a full quantum theory of target space using quantum theory of world-sheet. 
We propose that a proper analysis of the quantum  information in the would-sheet can be used to obtain the quantum states of target space.
To  properly  analyze the  information needed to construct  a quantum state of a system, we have to use  quantum estimation theory \cite{8a, 9a}. Now  a large  set of parameters in quantum estimation can be used to estimate the full quantum state of a system. This estimation of quantum state of the system by large  set of parameters is done using quantum state tomography \cite{qt12, qt14}. Thus, quantum state tomography has been  used to obtain quantum states of various different quantum   system  \cite{qt15, qt17, CS1}. It is  possible to perform quantum state tomography, even when the system has infinite degrees of freedom, such as a quantum field theory
\cite{qft1, qft2}. Several  interesting  techniques have been  developed  to perform quantum state tomography of such   system \cite{t1, t2, t4, t5}. 

The quantum state tomography has also been used to study the behavior of quantum states in loop quantum gravity \cite{lqg}. 
However, the quantum state tomography has not been used to analyze the behavior of quantum states of string theory. So,  we will perform  quantum state tomography of an open string. We would like to point out that in quantum state tomography for quantum mechanical systems, information obtained from  experiments is used to reconstruct  quantum state of the system. Motivated by this observation, we expect that in string theory, information obtained from  world-sheet should be used to construct quantum states of target space.  It may be noted that the that   string correlation functions have been used to analyze the behavior of world-sheet string states   \cite{core6, core7, corea1, corea2}. Now  as it is known that 
information about a  quantum system can obtain from correlation functions  \cite{core1, core2, core4, core5}, information theory has already been used to analyze the world-sheet string states.  We would like to also point out that the correction function of a quantum system can be  constructed from its tomogram \cite{tomoa1, tomoa2}, and this is another motivation to construct the  tomography in string theory. 

The connection between this quantum state of target space and classical geometry can be made using coherent states.
In fact, such a connection between   string coherent states and classical geometry has also been used in fuzzball proposal  \cite{4a,5a, 6a, 7a}. In this proposal, the entire region of space within  event horizon of a black hole is considered to be made up of  quantum state of strings.  This    fuzzball  states  is expressed as a wave functional in the full string theory, and not its  supergravity approximation. 
The fuzzball quantum state is   then  connected to classical geometry of a  black hole  using  string coherent states in target space, as these states can be approximated by  classical geometric solutions.   In fact,   it has been demonstrated that fuzzballs are   consistent with the  gravitational wave  observations done on  black holes \cite{7b, 7c}.  However, understanding of quantum nature of the string  in the fuzzball  is important to investigate  the black hole information paradox \cite{infor1, infor2}.  String coherent states are also important in the analysis of cosmic strings, which are  produced at the end of D3-D3 brane inflation   \cite{d1, d2}.  Furthermore, as  these cosmic  strings can be detected using     gravitational wave observation      \cite{b1, b2, b4, b5, b6}, it is important to understand their properties. 
Thus, it is important to understand the quantum states of target space coordinates, and this can be done by performing quantum states tomography using coherent states. 
We would like to point out that   quantum state tomography has been performed   using its coherent states \cite{co12, co14}.  Here we will generalize these results to string theory. 

To analyze such results in string theory, we need to first construct coherent states in string theory. The    coherent states of string have been constructed using  the DDF formalism \cite{ddf, ddf1}. The coherent states   in the Neveu-Schwarz sector has  been constructed using this formalism  \cite{ddf2}.  This has  been  extended to the Ramond sector by supersymmetric transformations in target space \cite{ddf2}.  
It is known that the DDF operators   satisfy the  oscillator algebra, and so they can be used to construct coherent string theoretical coherent states, which are analogous to the usual 
coherent states for quantum mechanical oscillator.
However, it is also possible to directly construct the string coherent states using the analogy of the original string algebra with the oscillator  algebra \cite{db,db12}.  So, in this paper, we will explicitly construct string coherent states. The  optical    coherent   states are important in    performed  quantum state tomography,   using quadrature operators  in optical phase space \cite{qd12, qd14, qd15, qd16}. In fact, optical  coherent states can be  constructed in quantum optics using  quadrature operators in optical phase space \cite{optics, optics4, optics5, optics1}. 
In this paper, we  will generalize the  quadrature operators  in optical phase space to string theory, and use them to perform quantum state tomography of an open  string.

\section {String Quadrature Operators}
\subsection{Bosonic String Theory}
In this section, we will review bosonic string theory, and  express it in a form where we can use the techniques from quantum optics  \cite{optics, optics4, optics5, optics1}. 
  The Polyakov action describes the world-sheet of bosonic  string theory, and it   can be written as 
\begin{eqnarray}
    \textbf{S} =  \frac{-T}{2} \int d\sigma ^2   \eta^{\alpha\beta} \partial_{\alpha}X_{\mu}  \partial_{\beta} X^{\mu}
\end{eqnarray}
where   the tension of the string  $T=1/\pi l^2_s$ is related to string length $l_s$. 
We can solve the above equation by applying Neumann boundary conditions for open strings.  
\begin{eqnarray}
\label{sol}
    X^{\mu}\left(  \tau,\sigma\right)=  x^{\mu}+l_s \tau p^\mu+ il_s\sum_{m\neq0}\frac{1}{m}{{\alpha}}^{\mu}_m e^{-im\tau}cos(m\sigma )
    \end{eqnarray}
As string  theory has a gauge degrees of freedom, we need to fix a guage before quantization it. In this paper, we will use the  light-cone gauge, where   the target  space coordinates     are defined as   $\{ X^+, X^-,X^i\}_{i=1}^{24} $, with $ X^+= ( X^0+X^{25}),$ and $  X^-= ( X^0-X^{25})$. In the light-cone gauge,  there are no oscillations in the $ X^+$ direction, and  the  oscillation in $X^-$ direction can be expressed in terms of other string  oscillations. Thus, we need to only consider string oscillations in the $\{  X^i\}_{i=1}^{24} $  direction to obtain information about the behavior of bosonic string theory in light-cone gauge.   
Now in the light-cone gauge,  the string algebra in these directions can be expressed as 
$
[\hat{\alpha}_m^i ,\hat{\alpha}_n^j] =m\eta^{ij} \delta_{m+n,0}
$,  which can also be written as
\begin{equation}
\label{comm}
[\hat{\alpha}_m^i ,\hat{\alpha}_{-m}^i] =m  \;\; ( \text{for}\; n =  -m   \;, \text{also},\; (\hat{\alpha}_m^i)^\dagger =  \hat{\alpha}_{-m}^i)
\end{equation}
with  $\{i,j\}$,  taking  values in $\{1, 2,3,..,24\}$.
We will suppress the index $i$ for simplicity, and express the   annihilation and creation operators as $\hat{\alpha}_m, \hat{\alpha}_{-m}$, and put them back at the end of the calculations. We can  now define the number operator $\hat{N}_m$  as the product of the annihilation and creation operators
\begin{equation}
\hat{N}_m =  (\hat{\alpha}_{-m} \cdot\hat{\alpha}_m)  \;  , \; \text{where}\;  m \; \ge 1
\end{equation}
We    define $|k\rangle$ as $\alpha_k|0\rangle $, and so  the eigenstates of the number operator satisfy
\begin{equation}
\hat{N}_m|k\rangle  = k_m|k\rangle
\end{equation}
where $k_m$ is the eigenvalue of $\hat{N}_m$. As a result, we have $\hat{N}_m (\hat{\alpha}_m|k\rangle) =(k -  m) (\hat{\alpha}_m|k\rangle)$ and $\hat{N}_m(\hat{\alpha}_{-m}|k\rangle) =(k +  m) (\hat{\alpha}_{-m}|k\rangle)$. Now, because  $(\hat{\alpha}_m|k\rangle $ and $(\hat{\alpha}_{-m}|k\rangle$ are eigenstates of $N_m$ with eigenvalues $ (k-m)$ and $(k+m)$, it follows that when $\hat{\alpha}_m $ and $\hat{\alpha}_{-m} $ operate on $ |k\rangle$, they decrease and increase $|k\rangle$ by '$m$' units.  This is an important point and a distinguishing factor of the world-sheet algebra. This  feature will be important in the construction of the quantum target space quadrature operators.
We can now write 
$
\hat{\alpha}_m |k\rangle  = C_k | k - m\rangle, $ and $
\hat{\alpha}_{-m}|k\rangle = D_k |k  +  m\rangle  $, where 
   $ C_k $ and  $D_k$ are proportionality constants. These proportionality constants can be  obtained as 
$(\langle k|\hat{\alpha}_{-m}) .(\hat{\alpha}_m |k\rangle )= \langle k| \hat{\alpha}_{-m} C_k |k-m\rangle=|C_k|^2$. 
The same product can also be calculated using 
$
(\langle k|\hat{\alpha}_{-m}) .(\hat{\alpha}_m |k\rangle ) =\langle k| N_m|k\rangle=k$. 
Now following the same procedure for $D_k$, we obtain the following proportionality constants $C_k=\sqrt{k}, $ and $D_k=\sqrt{k + m}$. 

\subsection{Construction of String Quadrature Operator  }
Now we will use techniques from quantum optics to construct and analyze the properties of   string quadrature operator  \cite{optics, optics4, optics5, optics1}. To  define  the string quadrature operators, we will first  define 
quadrature operators for  each string  would-sheet mode $m$. The quadrature operators are defined in optical phase space, such that they satisfy a commutation relation which is similar to the commutation relation satisfied by   the position and momentum operators of a quantum mechanical oscillator. Thus, the string   would-sheet mode quadrature operators would be expected to  satisfy  the canonical commutation relation
\begin{equation}
\label{commutator}
[ s_{\theta,m}, p_{\theta,m}] =i 
\end{equation}
These  would-sheet mode quadrature operators exist in a string analog of optical phase space. We observe that this commutation relation is satisfied if we define   would-sheet mode quadrature operators  in string world-sheet phase space (which is the string analog of optical phase space) 
\begin{eqnarray}
s_{\theta,m} &=&\frac{il_s \cos {m\sigma}}{m}\Bigg (\hat{\alpha}_m e^{-i(\theta+m\tau)} - \hat{\alpha}_{-m} e^{i(\theta+m\tau)}\Bigg)
\\
p_{\theta,m} &=&\frac{1}{2l_s\cos{ m\sigma}}\Bigg (\hat{\alpha}_m e^{-i(\theta+m\tau)} + \hat{\alpha}_{-m} e^{i(\theta+m\tau)}\Bigg)  
\end{eqnarray}
Here the string analog of the homodyne quadrature operators \cite{det1, det2, det4, det5} has been used to obtain information  encoded in the phase of the string modes. Thus, using the formalism of standard homodyne quadrature operators  \cite{det1, det2, det4, det5}, we have defined   $\theta$ to be  the  phase  associated with world-sheet homodyne quadrature operators, and so $0\leq \theta \leq 2\pi$.  Here in this string world-sheet phase space, $s_{\theta,m}, p_{\theta,m}$ act as coordinates and momentum, and  $\sigma, \tau$ act as parameters. 

Now from this perspective we can define fields on this world-sheet phase space, such that they would represent target space  coordinates. Thus, using these string   would-sheet mode quadrature operators,  we can define the targer space  quadrature operators  in quantum  target space  $ \hat{X}_{\theta},\hat{P}_{\theta}$, in the center of mass frame as 
\begin{eqnarray}
\hat{X}_{\theta} &=&il_s\sum_{m=1}^\infty\frac{\cos{ m\sigma}}{m}\Bigg (\hat{\alpha}_m e^{i(\theta+m\tau)}  - \hat{\alpha}_{-m} e^{-i(\theta+m\tau)}\Bigg)   =\sum_{m=1}^\infty  s_{\theta,m}
\\ 
\hat{P}_{\theta}&=& \sum_{m=1}^\infty\cos{m\sigma} \Bigg (\hat{\alpha}_m e^{i(\theta+m\tau)}  + \hat{\alpha}_{-m} e^{-i(\theta+m\tau)}\Bigg)   = 2\sum_{m=1}^\infty  \hat{p}_{\theta,m} \cos^2{m\sigma}
\end{eqnarray} 
It may be observed that  $\hat{P}_{\theta}$ can be viewed as the  momentum conjugate to $\hat{X}_{\theta}$. 
Here we had suppressed the index $i = \{1, 2,3,..,24\}$ in light-cone gauge, and if  write it explicitly, we  obtain the  quadrature operators for each of the target space coordinates $X^i_{\theta}$, and its momentum conjugate $ P^i_{\theta} $  in light-cone guage.  In general, we can perform the analysis in a general gauge and obtain quadrature operators $ X^\mu_{\theta},  P^\mu_{\theta} $.  Here different values of the world-sheet quadrature operators parameterized  by ${\theta}$ to construct the quantum state of the target space coordinates. This result generalized the classical relation between the target space coordinates and the world-sheet of a string. In classical geometry, the world-sheet of strings can be used as a probe for the target space, and information about the target space coordinates can be obtained using world-sheet of strings.  Here we have construed the quantum   target space using world-sheet phase space.

\subsection{Eigenstate of String Quadrature Operator}

In this section, we will derive an explicit expression for the eigenstates of the string quadrature operator. To do that, we  define eigenstates of the string   would-sheet mode quadrature operators as  $|s_{\theta,m}\rangle $ with the eigenvalue $ s_{\theta,m}$
\begin{equation}
\label{sthe}
\hat{s}_{\theta,m}  |s_{\theta,m} \rangle  = s_{\theta,m} |s_{\theta,m}\rangle    
\end{equation}
with $ s_{\theta,m} $ being the eigenvalue. Now the eigenstate of the string   quadrature operator $(\hat{X}_{\theta})$ can be written in terms of the eigenstates of the string  would-sheet mode   quadrature operators as  $|X_{\theta}\rangle=\prod_m |s_{\theta,m}\rangle$. Thus, this  quantum   target space    quadrature operator satisfies the following eigenvalue equation 
\begin{equation}
    \hat{X}_{\theta}|X_{\theta}\rangle= {X}_{\theta}|X_{\theta}\rangle
\end{equation}
where an explicit form for  $ {X}_{\theta}$ can be obtained from the eigenvalue equation of string  would-sheet mode quadrature operators  as 
\begin{align}
\langle k |\hat{X}_{\theta} |s_{\theta,m}\rangle\nonumber  &= s_{\theta,m}\bar{\psi}_{k} (s_{\theta,m})\\\nonumber
&=\langle k| \hat{\alpha}_m e^{-i(\theta+m\tau)}\frac{cos(m\sigma)}{m} -  \hat{\alpha}_{-_m} e^{i(\theta+m\tau)}\frac{cos(m\sigma)}{m}|s_{\theta,m}\rangle\\
&= [ \sqrt{k+m} e ^{-i(\theta+m\tau)} \bar{\psi}_{k+m} (s_{\theta,m})  -  \sqrt{k }e^{i(\theta+m\tau)} \bar{\psi}_{k-m} (s_{\theta,m}) ]
\end{align}
where  $
\langle k|s_{\theta,m}\rangle   =  \bar{\psi}_{k} (s_{\theta,m}), \,\, 
\langle k + m|s_{\theta,m}\rangle   =  \bar{\psi}_{k +m} (s_{\theta,m}) $ and $
\langle k - m|s_{\theta,m}\rangle   =   \bar{\psi}_{k - m} (s_{\theta,m})$
Now similarly, for  $
\psi_{k+m} (s_{\theta,m})  =  \langle s_{\theta,m} |k+m\rangle    
$, we can write 
\begin{equation}
\psi_{k+m}(s_{\theta,m})=\frac{e^{-i(\theta+m\tau)}}{\sqrt{k+m}\frac{cos(m\sigma)}{m}} [ s_{\theta,m} \psi_k(s_{\theta,m}) + \sqrt{k} e^{-i(\theta+m\tau)} \frac{cos(m\sigma)}{m}\psi_{k-m} (s_{\theta,m})] 
\end{equation}

Now we would like to find recurrence relations for this $\psi_m$, for different cases. We first observe that for  $k=0$, we can write 
\begin{equation}
\psi_{m}(s_{\theta,m})=\frac{e^{-i(\theta+m\tau)}}{\sqrt{m}\frac{cos(m\sigma)}{m}} [ s_{\theta,m} \psi_0(s_{\theta,m})]   
\end{equation}
Similarly, for any $k=n$, such that $n<m$,  the second term of recurrence will vanish (as $\psi_{n-m}=0$ for this case), and we can write 
\begin{equation}
\psi_{n+m}(s_{\theta,m})=\frac{e^{-i(\theta+m\tau)}}{\sqrt{n+m}
\frac{cos(m\sigma)}{m}} [ s_{\theta,m} \psi_n(s_{\theta,m})]  
\end{equation}
The second term does not vanish  $k\geq m$. 
So, for  $m=n$,  we have
\begin{equation}
\psi_{2m}(s_{\theta,m})=\frac{e^{-i(\theta+m\tau)}}{\sqrt{2m}\frac{cos(m\sigma)}{m}} [ s_{\theta,m} \psi_m(s_{\theta,m}) + \sqrt{m} e^{-i(\theta+m\tau)} \frac{cos(m\sigma)}{m}\psi_{0} (s_{\theta,m})]   
\end{equation}
and for    $k=l>m $, we have 
\begin{equation}
\psi_{l+m}(s_{\theta,m})=\frac{e^{-i(\theta+m\tau)}}{\sqrt{l+m}\frac{cos(m\sigma)}{m}} [ s_{\theta,m} \psi_l(s_{\theta,m}) + \sqrt{l} e^{-i(\theta+m\tau)} \frac{cos(m\sigma)}{m}\psi_{l-m} (s_{\theta,m})] 
\end{equation}
Here there are various different   recursion relations, each with  their own modified Hermite polynomials. We have obtained  the first three recursion relations to exemplify the procedure in Appendix,  and the any other  recursion relations  can be obtained  using the same  algorithm. These cases have been listed in Table 1. A consequence of this observation is that in string theory different states can be annihilated by a single annihilation operator.

\begin{table}[ht]
    \label{table}
    \begin{caption}
    {Values of '$m+k$' for different values of '$t$'}
    \end{caption}
    \centering
    \begin{tabular}{|l|c|c|c|r|}
    \hline \hline
     $k+m =$ & for $t = 1$ & for $t = 2$ & $\cdots$ & for $t = t$\\
    \hline
    $tm$ & $m$ & $2m$ & $\cdots$ & $tm$ \\
    \hline
    $tm+1$ & $m+1$ & $2m+1$ & $\cdots$ & $tm+1$ \\
    \hline
    $tm+2$ & $m+2$ & $2m+2$ & $\cdots$ & $tm+2$\\
    \hline
    \vdots & \vdots & \vdots & \vdots & \vdots\\
    \hline
    $tm+(m-1)$ &  $2m-1$ &  $3m-1$ & $\cdots$ & $tm+(m-1)$\\
    \hline
    \end{tabular}
\end{table}
Using these recurrence relation,  we can calculate expressions of $\psi_0,\psi_1,...\psi_{m-1}$, which can then be used to obtain   $m^{th}$  would-sheet mode wave function $\psi_m$. In order to do this, we start by writing the annihilation operator $\hat{\alpha}_m$ in terms of the quadrature operators $(\hat{s}_{\theta,m},\hat{p}_{\theta,m})$  as 
\begin{equation}\label{a}
    \hat{\alpha}_m=\frac{1}{2}e^{-i(m\tau+\theta)}\Big( {\hat{s_{\theta,m}}}+\frac{2 {l_s}^2cos^2(m\theta)}{m i}\hat{p_{\theta,m}}\Big)
\end{equation}
We can use the annihilation operator to annihilate $\psi_0$ as $\hat{\alpha}_m\psi_0=\langle {s_{\theta,m}}|\hat{\alpha}_m|0\rangle=0$. 
As   would-sheet mode quadrature operators  satisfy the commutation algebra of position and momentum operators, we can express $\hat{p}_{\theta,m}$ as $-i d/d \hat{s}_{\theta,m}$, and  write 
\begin{eqnarray}
 {\Big(\hat{s}_{\theta,m}+\frac{2 l_s^2 cos^2(m\sigma)}{m}\frac{d}{d\hat{s}_{\theta,m}}\Big)}\psi_0=0
\end{eqnarray}
As a result, we get an explicit expression for $\psi_0$ (with $A$ as the normalization constant), 
\begin{eqnarray}
\psi_0=A e^{-\frac{m {s_{\theta,m}}^2}{4{l_s}^2 cos^2(m\sigma)}}, &\text{with} &     A = \Big[\frac{m}{2\pi l_s^2\cos^2(m\sigma)}\Big]^{\frac{1}{4}}
\end{eqnarray}
Similarly, because $\hat{\alpha}_m|z\rangle=0$,  where $z=(0,1....m-1)$, we have $\psi_z=\psi_0$. This behavior of $\psi_m$ can be clearly seen in Fig. 1. For instance, for $m=1$ case, all the wave functions will be different, since $m=1$ replicates the usual quantum mechanical behavior. However, for $m=2$, $\psi_0=\psi_1$. We have plotted $\psi_4$ 
to illustrate this behavior.

\begin{figure}[h!]

            \begin{center}
            $%
            \begin{array}{cccc}
            \includegraphics[width=60 mm]{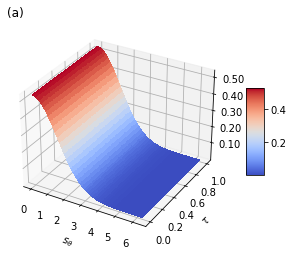} \includegraphics[width=60 mm]{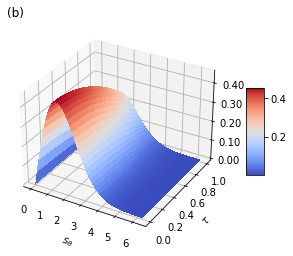}\\ \includegraphics[width=60 mm]{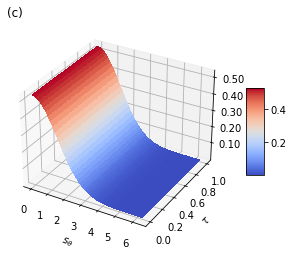} \includegraphics[width=60 mm]{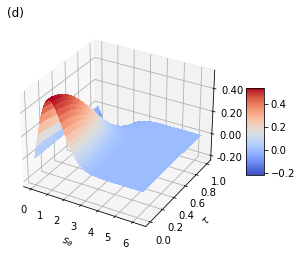}
            
            \end{array}%
            $
            \caption{The plots show the behavior of $\psi_m$ for the special case of $m=1,\sigma=0,\theta=0$ as a function of $s_{\theta,m}$ and time $\tau$. (a) Ground state wavefunction $\psi_0$ which preserves its Gaussian shape through time, whereas (b) is the first excited state $\psi_1$ which oscillates in time sinusoidally. (c) $\psi_0$ for $m=2$ and (d) $\psi_4$ for m=2.}
            \end{center}
\label{figure}            
\end{figure}


\section{String State Tomography}
\subsection{ String Coherent States}
The  DDF formalism has been used to construct string   coherent   states    \cite{ddf, ddf1}. The DDF formalism has   been used in both    Neveu-Schwarz sector  \cite{ddf2} and    \cite{ddf2}.  The string coherent states have also been constructed using the  analogy of the original string algebra with the oscillator  algebra \cite{db,db12}.     It may be noted that the optical  quadrature operators have been used to obtain coherent states in quantum optics  \cite{optics, optics4, optics5, optics1}. Motivated by this construction,   we will construct the string coherent states in target space using  string quadrature operators obtained in the previous section. 
Now to construct string coherent states, we will first construct the 
coherent states for a single would-sheet mode. Thus, we can define the string  would-sheet mode coherent state as states which  most closely resembling  behavior of a classical string  oscillatory modes. So, the string  would-sheet mode coherent state $|\varphi_m\rangle$ can be expressed as an eigenstate of $\hat\alpha_m$, with eigenvalue $\varphi_m$
\begin{equation}
\label{STS}
\hat\alpha_m |\varphi_m\rangle = \varphi_m|\varphi_m\rangle    
\end{equation}
where        $ |\varphi_m\rangle$ satisfying  $ \langle\varphi_m|\varphi_m\rangle = 1 $. In order to derive an explicit expression for  string coherent state $ |\varphi_m\rangle$  we expand  in terms of  $ |k\rangle$ string states as 
\begin{equation}
\label{STS8}
|\varphi_m\rangle =  \sum_{k=0}^{\infty} |k\rangle \langle k|\varphi_m\rangle
\end{equation}
Here we can use $\langle k|\hat\alpha_m | \varphi_m\rangle = \varphi_m \langle k|\varphi_m\rangle$ to obtain 
$\langle k+m|\varphi_m\rangle = {\varphi_m} \langle k|\varphi_m\rangle   /{\sqrt{k+m}} $.  Replace $k$ with $k - m$, we  obtain 
$
\langle k|\varphi_m\rangle ={\varphi_m}\langle k-m|\varphi_m\rangle    /{\sqrt{k}} $, and by repeating this    process, we also obtain 
$
\langle k-m|\varphi_m\rangle = {\varphi_m} \langle k-2m|\varphi_m\rangle/{\sqrt{k-m}}$. Using these expression, we can write 
$
\langle k|\varphi_m\rangle = {\varphi_m^2} \langle k-2m|\varphi_m\rangle/{\sqrt{k(k-m)}}
$. Now 
repeating this procedure $n$ times, we obtain a general expression for $\langle k|\varphi_m\rangle$ as 
\begin{equation}
\langle k|\varphi_m\rangle =\frac{\varphi_m^n}{\sqrt{k(k-m)(k-2m)...(k-(n-1)m)}} \langle k-nm|\varphi_m\rangle   
\end{equation}
Now for $k\geq m$, and    $k-nm =b$, with $b=(0,1,2...m-1), \, n = (0, 1, 2, 3...)$,   we can write the expression for $\langle k|\varphi_m\rangle$ as 
\begin{equation} 
\label{STS5}
\langle k|\varphi_m\rangle =\frac{\varphi_m^{n}}{\sqrt{k(k-m)(k-2m)...m)}} \langle b|\varphi_m\rangle    
\end{equation}

We can obtain the expression for  $ \langle b |\varphi_m\rangle $ using the string displacement operator $D(\varphi_m)$) and the  normalization condition. We define the string  would-sheet mode  displacement operator  as the operator which  generates a  string  would-sheet mode coherent states from the vacuum state, $|\varphi_m\rangle = D(\varphi_m) |0\rangle$. Thus, we can write the explicit expression for string  would-sheet mode displacement operator as 
\begin{equation}
\hat{D}(\varphi_m)= e^{{-\frac{|\varphi_m|^2}{2}}} e^{{\varphi_m \hat\alpha_{-m}}} e^{{\varphi_m^* \hat\alpha_m}}
\end{equation} 
Now to obtain the expression for $ \langle b|\varphi_m\rangle $, we observe that 
\begin{align}
\langle b|\varphi_m\rangle &= \langle b|  D(\varphi_m) |0\rangle =e^{{-\frac{|\varphi_m|^2}{2}}} \langle b| e^{{\varphi_m \hat\alpha_{-m}}} e^{{\varphi_m^* \hat\alpha_m}}  |0\rangle
\nonumber  
\\ 
&=   e^{{-\frac{|\varphi_m|^2}{2}}} \langle b| ( 1+\varphi_m \hat\alpha_{-m}+....)  ( 1+\varphi_m^*\hat\alpha_{m}+...) |0\rangle   
\end{align}
Thus,   using $ \hat\alpha_{m} |0\rangle = 0$, we obtain 
$
\langle b|\varphi_m\rangle = e^{{-{|\varphi_m|^2}/{2}}}  \delta_{0,b}   
$, which vanishes for $b \neq 0$, and so, we can write 
\begin{equation}
\label{STS6}
\langle 0|\varphi_m\rangle = e^{{-\frac{|\varphi_m|^2}{2}}}    
\end{equation}
In the  general expression for $\langle k|\varphi_m\rangle$, we observe that $\langle b |\varphi_m\rangle =0$ for $b \neq 0$, and  we obtain a non-vanishing expression only for  $k=nm$. So, 
using the expression for $\langle b|\varphi_m\rangle$ in the general expression for $\langle k|\varphi_m\rangle$, we obtain 
\begin{equation}
\label{STS7}
\langle k|\varphi_m\rangle = \frac{\varphi_m^{n} e^{{-\frac{|\varphi_m|^2}{2}}} }{\sqrt{k(k-m)(k-2m)...(k-(n-1)m)}}    
\end{equation}
Using this expression for $\langle k|\varphi_m\rangle$ (with $k =nm$) in the general expression for string  would-sheet mode coherent states, we can write an explicit expression for a string  would-sheet mode coherent state as  
\begin{equation}\label{60}
|\varphi_m\rangle =\sum_{n=0}^{\infty}  \frac{\varphi_m^{n} e^{{-\frac{|\varphi_m|^2}{2}}} }{\sqrt{k(k-m)(k-2m)...(m)}}  |k\rangle
\end{equation} 
By repeating this procedure for different string modes, we can obtain string  would-sheet mode coherent states for different string  modes.
Using these string  would-sheet mode coherent states, a coherent state for strings in target space  can be expressed as   
\begin{equation}
 |\Phi\rangle 
 =\prod_m |\varphi_m\rangle
\end{equation}
Thus, we have generalized the construction of coherent states to string theory using coherent states for each of the string modes. 
It may be noted that even though string coherent states have been obtained before, this is the first time that they have been obtained using string analogs of the quadrature operators.  We can construct a coherent state for each of the target space coordinates by explicitly introducing the index. Thus, in light-cone gauge,  we can repeat this procedure to construct a full set of coherent states for each of the target space coordinates in the light-cone gauge $\{ {|\Phi\rangle^i  } \}$. These can be constructed from string  would-sheet mode coherent states, $ |\Phi\rangle^i =\prod_m |\varphi^i_m\rangle$, where we can define the string  would-sheet mode coherent states for various target space coordinates  in light-cone gauge using  
$\hat\alpha_m^i |\varphi_m^i\rangle = 
\varphi_m^i|\varphi_m^i\rangle  $. Thus, the world-sheet coherent states can be used to construct coherent states for target space, which can be approximated by classical geometric solutions.

\subsection{Quantum State Tomography} 
In the previous section, we obtained string  coherent states in target space. As coherent states can be   used to perform    quantum state tomography for a system  \cite{co12, co14}, we will use the string coherent states obtained in the previous section to perform quantum state tomography of bosonic strings. To do this we will first expand the eigenstates of a string  would-sheet mode quadrature operator as  $
|s_{\theta,m}\rangle= \sum_{t=0}^{\infty} |t \rangle \langle t |s_{\theta,m}\rangle
$, with 
  $|t \rangle$ the string  excited state. Thus, using   $\langle t |s_{\theta,m}\rangle=\bar{\psi}_t (s_{\theta,m})$, we can write 
\begin{align}\label{63}
|s_{\theta,m}\rangle&= \sum_{t=0}^{\infty} |t \rangle \bar{\psi}_t(s_{\theta,m})\\\nonumber
 &=\bar{\psi}_0\sum_{t=0}^{m-1}|t\rangle+
\sum_{t=m}^{\infty} \bar{\psi}_0(s_{\theta,m}) J_{t} (s_{\theta,m}) e^{it(\theta+m\tau)} |t\rangle 
\end{align}
Using this expression for a string  would-sheet mode quadrature  operator,  we can write the eigenstates of a string quadrature operator as   
\begin{equation}
|X_{\theta}\rangle =  \prod_m |s_{\theta,m}\rangle. 
\end{equation}
Now from  this expression for the eigenstates of a string quadrature operator  in quantum target space $|X_{\theta}\rangle $, we can define the quantum state tomogram for a string as 
$\Omega(S_{\theta},\theta)$, where 
\begin{equation}
\Omega( X_{\theta}, \theta) = \langle X_{\theta} |\hat{\rho}|X_{\theta} \rangle    
\end{equation}
with $\hat{\rho}$ as the string  density matrix for the given system. Here we have also used   the normalization condition  
\begin{equation}
\int \Omega( X_{\theta} , \theta) dX_{\theta}  =  1    
\end{equation}
It may be noted that in analogy with quantum optics \cite{optics, optics4, optics5, optics1},  the tomogram of the pure  open string state in target space is represented by the density matrix $\hat{\rho} =|\Phi\rangle \langle\Phi|$ can be written as $\Omega( X_{\theta} , \theta) = | \langle X_{\theta}| \Phi\rangle|^2$. To evaluate this tomogram for an open string, we need to express it in terms of   tomogram for would-sheet string modes, which can be written as   $ \Omega( s_{\theta,m} , \theta_m) = | \langle s_{\theta,m}| \varphi_m\rangle|^2  $.  
Thus, to perform quantum state tomography of an open  string, we   evaluate $ \langle s_{\theta,m}\vert \varphi_m \rangle$, and 
 write a tomogram for world-sheet string modes as 
\begin{align}
\label{SMCS}
 |\langle s_{\theta,m}\vert \varphi_m \rangle|^2 =& \Bigg\lvert
 \sum_{t=0}^{\infty}  \sum_{n=0}^{\infty} 
    \bigg[ \frac{J_{tm} (s_{\theta,m}) e^{-it(\theta+m\tau)} \psi_0(s_{\theta,m}) \varphi_m^n e^{-\frac{|\varphi_m|^2}{2}}}{\sqrt{nm((nm)-m)((nm)-2m)...m}} \langle tm|nm \rangle\bigg] \Bigg\rvert^2     \nonumber\\
 =& \Bigg\lvert  \sum_{n=0}^{\infty}   
     \frac{J_{nm} (s_{\theta,m}) e^{-in(\theta+m\tau)} \psi_0(s_{\theta,m}) \varphi_m^n e^{-\frac{|\varphi_m|^2}{2}}}{\sqrt{nm((nm)-m)((nm)-2m)...m}}    \Bigg\rvert^2  
\end{align}
Now  using this expression for  tomogram for would-sheet string modes, we can write the tomogram for the target space as 
\begin{eqnarray}
\label{Tom}
   \lvert  \langle     X_{\theta}|\Phi\rangle\rvert^2  &=& \Big|\prod_m \langle s_{\theta,m}\vert \varphi_m \rangle\Big|^2
 \\&=&  \prod_{m} \Bigg\lvert \sum_{n=0}^{\infty}  \bigg[ \frac{J_{nm} (s_{\theta,m}) e^{-in(\theta+m\tau)} \psi_0(s_{\theta,m}) \varphi_m^n e^{-\frac{|\varphi_m|^2}{2}}}{\sqrt{nm((nm)-m)((nm)-2m)...m}} \bigg]\Bigg\rvert^2   \nonumber
\end{eqnarray}

Thus, we can perform the quantum state tomography in a quantum  target space, using this expression in terms of would-sheet modes. Here if we do not suppress the index, and repeat this procedure we can express all the information needed to construct a quantum  target space using   $\{\Omega^i( X^i_{\theta} , \theta)\} $, where $ {\Omega^i( X^i_{\theta} , \theta)= | \langle X^i_{\theta}| \Phi\rangle|^2}$, with $i = \{1, 2,3,..,24\}$ in light-cone coordinates. We plot the tomogram for 
target space in Fig. 2, Fig. 3 and Fig. 4 with various parameter choices. In Fig. 2, we see the variation of the tomogram as we add modes to the system. It is clear that the addition of modes centralizes the tomogram. This is evident from Eq. \ref{Tom}, when we take the product of multiple modes, there is constructive interference in the central region but destructive interference elsewhere. Fig. 3 is the variation of the tomogram with the time-like coordinate $\tau$. Changing $\tau$ simply translates the tomogram and this is clear from Eq. \ref{Tom}. Lastly, in Fig. 4 we investigate the variation of the tomogram with different values of $\sigma$, the space-like coordinate on the world sheet as can be seen in Eq. \ref{sol}. Due to the structure of the modified Hermite polynomials $J_{(t)m}$, we expect that the tomogram is undefined when $m\sigma=\frac{\pi}{2}$ and thus we vary $\sigma$ through odd divisions of $\pi$. From this it is clear why the tomogram squishes for certain values of $\sigma$.


\begin{figure}[h!]\label{c}
            \begin{center}
            $%
            \begin{array}{cccc}
            \includegraphics[width=70 mm]{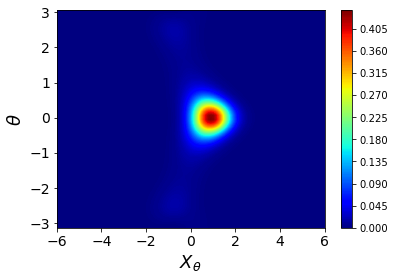}
            \includegraphics[width=70 mm]{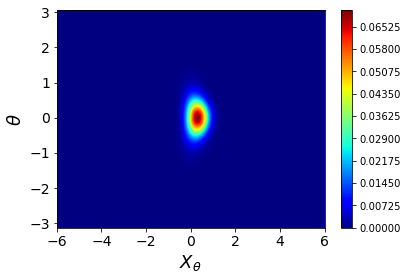}\\
            \includegraphics[width=70 mm]{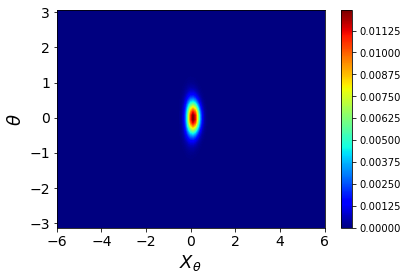}
            \includegraphics[width=70 mm]{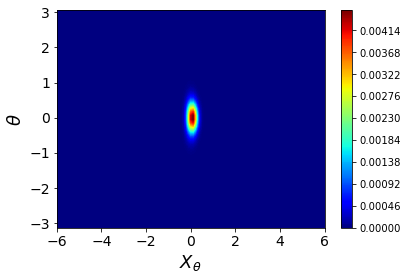}
            \end{array}%
            $%
            \caption{Contour plot of the tomogram $\Omega(X_{\theta,m},\theta)$, where $N=20$, $\tau=0$, $\sigma = 0$ and $M=\{3,5,8,10\}$. Figures are enumerated from left to right}
            \end{center}
\label{fig2}            
\end{figure}

\begin{figure}[h!]\label{c}
            \begin{center}
            $%
            \begin{array}{cccc}
            \includegraphics[width=70 mm]{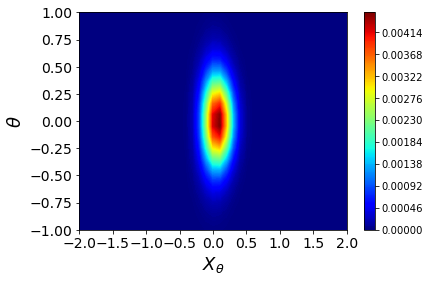}
            \includegraphics[width=70 mm]{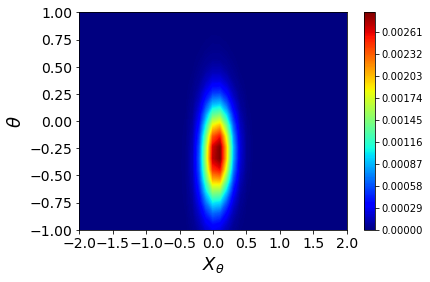}
            \end{array}%
            $%
            \caption{Contour plot of the tomogram $\Omega(X_{\theta,m},\theta)$, where $M=10$, $N=20$. For $\sigma =0$, it shows the time evolution $\tau = \{0,0.1\}$. It simply translates as time progresses and this is clear from Eq. \ref{Tom}}
            \end{center}
\label{fig4}            
\end{figure}

\begin{figure}[h!]\label{c}
            \begin{center}
            $%
            \begin{array}{cccc}
            \includegraphics[width=50 mm]{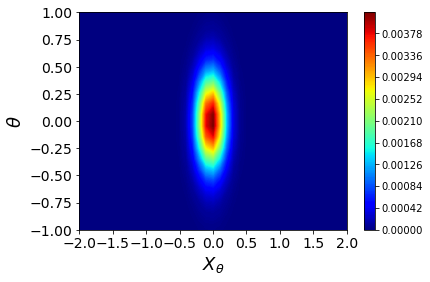}
            \includegraphics[width=50 mm]{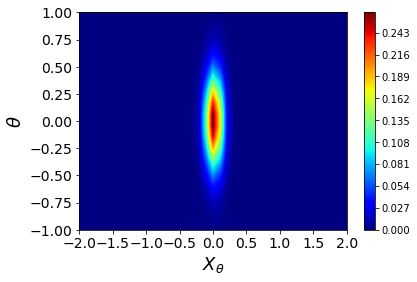}
            \includegraphics[width=50 mm]{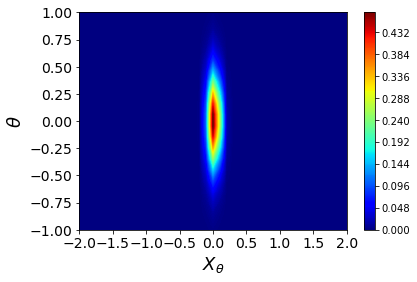}\\
            \includegraphics[width=50 mm]{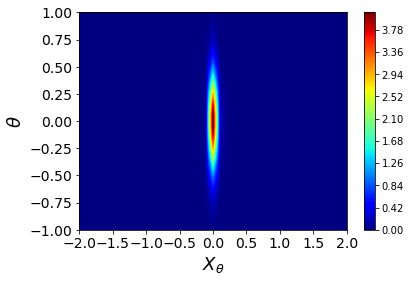}
            \includegraphics[width=50 mm]{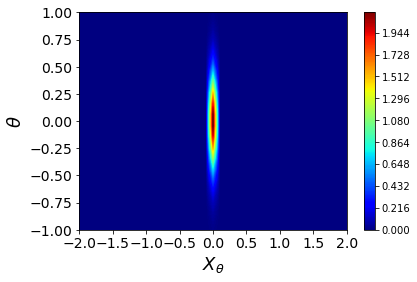}
            \includegraphics[width=50 mm]{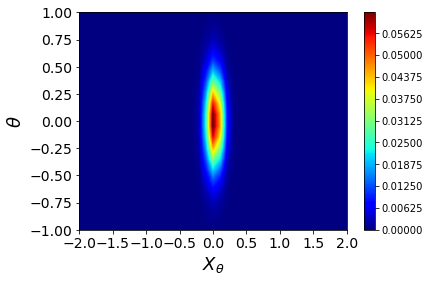}\\
            \includegraphics[width=50 mm]{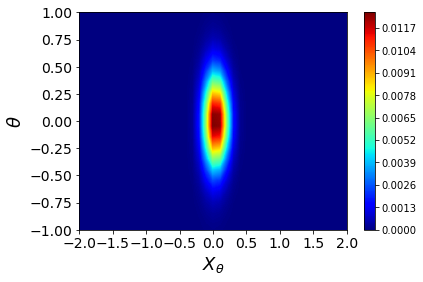}
            \includegraphics[width=50 mm]{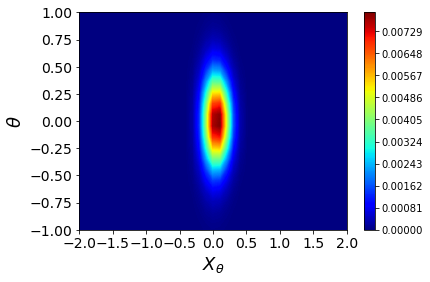}
            \includegraphics[width=50 mm]{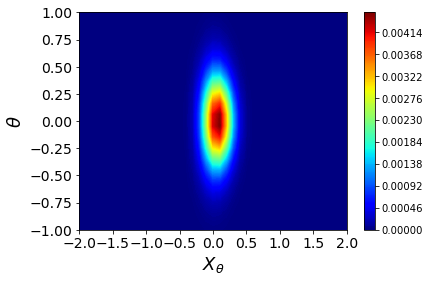}
            \end{array}%
            $%
            \caption{Contour plot of the tomogram $\Omega(X_{\theta,m},\theta)$, where $M=10$, $N=20$. For $\tau=0$ and $\sigma = \{\pi,\pi/3,\pi/7,\pi/11,\pi/19,\pi/25,\pi/37,\pi/49,0\}$. Figures are enumerated from left to right}
            \end{center}
\label{fig3}            
\end{figure}

            %

\section{Conclusion}
It is known that  the quantum state tomography contains all the information needed to construct a given quantum state. This motivated the construction of quantum state tomography in string theory. 
To perform  quantum state tomography of an open string, we  constructed   quadrature operators  for an open string in quantum   target space. These string  quadrature operators were constructed using  quadrature operators for different world-sheet  modes of an open string.  We also defined a suitable string  displacement operator which would convert a vacuum state into a string   world-sheet mode coherent state. These string  world-sheet mode coherent states were then used to construct the coherent state for an open string. 
We used these coherent states along with the string quadrature operators   to perform quantum state tomography of an open string.  
We would like to point out that we performed  quantum state tomography on a flat target space. However, it would be interesting to generalize these results to open string coupled to gravity.   Furthermore, it would be interesting to perform the quantum state tomography for string states representing a fuzzball and a cosmic string. We expect to obtain first order corrections to the classical behavior of such states using quantum tomography. These results can then be used to analyze black hole information paradox \cite{infor1, infor2}. To perform this analysis it is important to analyze the quantum tomogram for strings in curved space-time. 

It would be interesting to generalize these results to closed strings. We expect that the quadrature operators  for an closed string could also be constructed using the quadrature operators  for different modes of a closed string.  {We expect that we will need to construct too the quadrature operators, which would correspond to right and left movers.} We can also obtain coherent states for a closed string using the same algorithm. These closed string coherent states along with the corresponding quadrature operators can be used to perform quantum state tomography for closed strings. 
It would also be interesting to  generalize these results to thermal string states. We can construct thermal coherent states for those thermal states   \cite{q16}. This can again be done by defining thermal coherent string  would-sheet mode states, and then using them to construct the thermal coherent states for strings.  These thermal coherent states can then be used to perform  quantum tomography for such thermal string states. This can be done for both open and closed string states.  

We would like to point out that coherent states of strings are also important in the context of AdS/CFT correspondence. This is because 
the micro-states of geometric objects like black holes can be analyzed using   string states \cite{1a,2a}. This is because the micro-states of an AdS black hole can  be obtained from the micro-states of the  conformal field theory dual to it, using the AdS/CFT correspondence  \cite{ads1, ads2}. It is possible to obtain information about  correlation functions of  a conformal field theory by  using the Skenderis-van-Rees prescription  \cite{q12, q14}.  This prescription is based on  AdS/CFT correspondence, and in it   the   initial and final states are represented by   coherent states in AdS spacetime  \cite{q15}. The   Skenderis-van-Rees prescription  has been generalized to conformal field theories at  finite temperature, which are dual to a black hole in AdS \cite{q16}.
Thus, it is important to investigate  coherent states in AdS to understand the conformal field theories dual to black holes. 
It would be interesting to investigate this correspondence beyond supergravity approximation. This can be done by analyzing a possible duality between quantum state tomography of  target space in AdS with the quantum state tomography of the boundary conformal field theory.

\section*{Appendix}
Here we explicitly derive recurrence relation for three different cases. This will be done by   calculating  the complete set of these modified Hermite polynomials. 
\vspace{2ex}
\begin{itemize}
    \item \textbf{Case $1$: k+m=tm}
    We start from 
\begin{eqnarray}
 \psi_{k+m}(s_{\theta,m})&=&\frac{e^{-i(\theta+m\tau)}}{\sqrt{k+m}\frac{cos(m\sigma)}{m}} [ s_{\theta,m} \psi_k(s_{\theta,m}) + \nonumber\\
 &&\sqrt{k} e^{-i(\theta+m\tau)} \psi_{k-m} \frac{cos(m\sigma)}{m}(s_{\theta,m})] \end{eqnarray}
For $t=1$, we obtain
{\begin{equation}
\psi_{m}(s_{\theta,m})=\frac{e^{-i(\theta+m\tau)}}{\sqrt{m}\frac{cos(m\sigma)}{m}} [ s_{\theta,m} \psi_0(s_{\theta,m})]      
\end{equation}}
and for $t=2$, we obtain  is
{\begin{align}
\psi_{2m}(s_{\theta,m})&=\frac{e^{-i(\theta+m\tau)}}{\sqrt{2m}\frac{cos(m\sigma)}{m}} [ s_{\theta,m} \psi_m(s_{\theta,m}) + \sqrt{m} e^{-i(\theta+m\tau)} \frac{cos(m\sigma)}{m}\psi_{0} (s_{\theta,m})]\nonumber\\
&=\frac{e^{-2i(\theta+m\tau)}}{\sqrt{2m}\frac{cos(m\sigma)}{m}}[ \frac{1}{\sqrt{m}\frac{cos(m\sigma)}{m}} s_{\theta,m} s_{\theta,m} + \sqrt{m}\frac{cos(m\sigma)}{m}] \psi_0(s_{\theta,m}) 
\end{align}}
In general, $\psi_{tm}$ can be written as follows
{\begin{equation}
\psi_{tm}(s_{\theta,m})=e^{-ti(\theta+m\tau)} J_{(t)m} (s_{\theta,m}) \psi_0 (s_{\theta,m})
\end{equation}}
where $J_{(t)m} $ is a new polynomial defined by the following recurrence relation;
\begin{eqnarray}
J_{(t+1)m} (s_{\theta,m}) &=&\frac{1}{\sqrt{(t+1)m}\frac{cos(m\sigma)}{m}} [  s_{\theta,m} J_{(t)m} (s_{\theta,m}) \nonumber\\
&&+\sqrt{(tm)}\frac{cos(m\sigma)}{m} J_{(t-1)m}(s_{\theta,m})]
\end{eqnarray}
with  $J_m=\frac{1}{\sqrt{m}\frac{cos(m\sigma)}{m}}(s_{\theta,m})$ and $J_0 =1$.
\item \textbf{Case $2$: k+m=tm+1}
Here we use 
\begin{align}
\psi_{k+m}( s_{\theta,m}) = \psi_{tm+1} ( s_{\theta,m}) = \frac{e^{-i(\theta+m\tau)} }{ \sqrt{tm+1}\frac{cos(m\sigma)}{m} } [  ( s_{\theta,m}) \psi_{(t-1)m+1 } \\+
\sqrt{(t-1)m+1} e^{-i(\theta+m\tau)}\frac{cos(m\sigma)}{m}  \psi_{(t-2)m+1} ]  
\end{align}
We also use 
\begin{equation}
\psi_{tm+1}(s_{\theta,m}) =e^{-ti(\theta+m\tau)}J_{(tm)+1}(s_{\theta,m}) \psi_{1}(s_{\theta,m})
\end{equation}

Now $J_{(t-1)m+1}(s_{\theta,m})$ is a new polynomial defined by the following recurrence relation;
\begin{eqnarray}
J_{(t+1)m+1}(s_{\theta,m}) &=& \frac{e^{-ti(\theta+m\tau)}}{\sqrt{(t+1)m+1}\frac{cos(m\sigma)}{m} } [ s_{\theta,m} J_{m(t)+1}(s_{\theta,m}) \nonumber\\&&+ \sqrt{m(t)+1}\frac{cos(m\sigma)}{m}  J_{m(t-1)+1}(s_{\theta,m})]
\end{eqnarray}
with  $J_{m+1}=\frac{1}{\sqrt{m+1}\frac{cos(m\sigma)}{m}}(s_{\theta,m})$ and $J_1 =1$. 
 {Similarly, we can write the polynomial expressions for $k+m = tm+2$ and other such expressions. Hence, we can directly derive the expression for $k+m=tm + (m-1)$.}
\item \textbf{Case $m-1$: k+m=tm+(m-1)}
Here we use 
\begin{eqnarray}
\psi_{k+m}(s_{\theta,m})&=& \psi_{(t+1)m-1}(s_{\theta,m}) \nonumber\\
&=&\frac{e^{-i(\theta+m\tau)}}{\sqrt{{tm}+{(m-1)}}\frac{cos(m\sigma)}{m} }\bigg[s_{\theta,m} \psi_{(t-1)m+(m-1)}(s_{\theta,m})
\nonumber\\+&&
\sqrt{(t-1)m+(m-1)}\frac{cos(m\sigma)}{m} e^{-i(\theta+m\tau)} \psi_{(t-2)m+(m-1)}(s_{\theta,m})\bigg] \nonumber\\
&=& e^{-ti(\theta+m\tau)} J_{(t+1)m-1}(s_{\theta,m}) \psi_{m-1}(s_{\theta,m})   
\end{eqnarray}

Now $J_{(t+1)m-1}$ is a new polynomial defined by the following recurrence relation;
\begin{eqnarray}
J_{(t+2)m-1}(s_{\theta,m})& =&\frac{1}{\sqrt{(t+2)m-1}\frac{cos(m\sigma)}{m}} [  s_{\theta,m} J_{(t+1)m-1}(s_{\theta,m}) \nonumber \\
&&+\sqrt{{(t+1)m-1}} \frac{cos(m\sigma)}{m}J_{(tm)-1}(s_{\theta,m})]
\end{eqnarray}
with $J_{2m-1}=\frac{1}{\sqrt{2m-1}\frac{cos(m\sigma)}{m}}(s_{\theta,m})$ and $J_{m-1} =1$.
\end{itemize}

\end{document}